\documentclass[conference,compsoc]{IEEEtran}

\ifCLASSOPTIONcompsoc
  \usepackage[nocompress]{cite}
\else
  \usepackage{cite}
\fi

\usepackage[T1]{fontenc}
\usepackage[utf8]{inputenc}
\usepackage{microtype}
\usepackage{amsmath,amssymb}
\usepackage{graphicx}
\usepackage{xcolor}
\usepackage{url}
\usepackage{hyperref}
\usepackage{booktabs}
\usepackage{longtable}
\usepackage{array}
\usepackage{enumitem}
\usepackage{csquotes}
\usepackage{float}
\usepackage{caption}
\usepackage{booktabs}
\usepackage{tabularx}
\usepackage{array}
\usepackage{float}

\hypersetup{
  colorlinks=true,
  linkcolor=blue,
  citecolor=blue,
  urlcolor=blue
}

\definecolor{revisiongreen}{RGB}{0,105,0}

\begin{document}

\title{Digital Security Under Compound Threats: Surveillance, Threat Reasoning, and Habituation Among North Korean Defectors in China}

\author{
\IEEEauthorblockN{Yibo Meng\IEEEauthorrefmark{1},
Bingyi Liu\IEEEauthorrefmark{2},
Lionel Z. Wang\IEEEauthorrefmark{3},
and Hongkun Yang\IEEEauthorrefmark{4}}
\IEEEauthorblockA{\IEEEauthorrefmark{1}Cornell University, mengyb22@tsinghua.org.cn}
\IEEEauthorblockA{\IEEEauthorrefmark{2}University of Michigan--Ann Arbor, bingyi@umich.edu}
\IEEEauthorblockA{\IEEEauthorrefmark{3}Nanyang Technological University, lionel-z.wang@connect.polyu.hk}
\IEEEauthorblockA{\IEEEauthorrefmark{4}Ocean University of China, Hongkunyang16@gmail.com}
}

\maketitle

\begin{abstract}

North Korean defectors residing in China face persistent uncertainty regarding digital security and privacy. Through semi-structured interviews with 19 defectors, we examine how individuals understand digital threats and adapt their security behaviors. We identify three interrelated conditions: transnational surveillance pressure from North Korea, identity-linked digital infrastructure and intermediary dependency in China, and threat reasoning shaped by prior surveillance schemas. Participants developed uneven protective practices and followed two trajectories: adaptive habituation driven by knowledge acquisition and erosive habituation driven by the prolonged absence of adverse events. Our findings show how compound threats, attribution uncertainty, and constrained agency shape security reasoning among undocumented users.

\end{abstract}

\IEEEpeerreviewmaketitle

\section{Introduction}

North Korea operates one of the most pervasive state surveillance systems in the world, monitoring citizens through neighborhood-level reporting networks, collective punishment, and strict control over information and movement~\cite{warf2015hermit,lankov2014north}. Each year, thousands flee these conditions, and China is one of the primary destinations~\cite{lankov2004north,chang2008migration}. Those who cross are commonly referred to as North Korean defectors (NKDs). Because the Chinese government has largely treated them as illegal economic migrants rather than refugees, most NKDs remain undocumented and vulnerable to detention and forced repatriation~\cite{cho2013systemizing,beck2007perilous}.

For undocumented NKDs, mobile phones and platform accounts are not optional conveniences but the infrastructure through which people contact family, coordinate with employers and brokers, seek work, arrange transportation, and learn about community support. The same infrastructure exposes them: a SIM card may require identity documents they cannot safely present; a platform account may link an improvised identity to a device, phone number, or movement pattern; a phone may contain call records, contacts, and location traces that become evidence if inspected; and a device provided by an employer creates both access and a channel of dependence and control. Digital participation is therefore an ongoing security and privacy problem in which communication, identity, mobility, and survival are tightly entangled.

Human-centered security research has long shown that security mechanisms fail when they do not fit users' goals, constraints, mental models, and social environments~\cite{whitten1999johnny,adams1999users,wash2010folk,krombholz2019if}, and that at-risk users face amplified likelihood and severity of digital harm~\cite{warford2022sok}. Prior work on refugees, migrants, journalists, and activists shows protective practices diverge from expert recommendations when users must prioritize access, livelihood, or safety~\cite{simko2018computer,mcgregor2017weakest,sambasivan2018privacy,tran2024security}. Much usable-security guidance assumes that users can identify relevant threats, control their devices and accounts, obtain usable advice, and update behavior through feedback~\cite{wash2010folk,krombholz2019if,herley2009so,redmiles2020comprehensive}. These assumptions break down when legal identity itself creates risk, infrastructure is identity-linked, devices are controlled by third parties, and multiple actors are difficult to distinguish.

NKDs in China exemplify these limits. China is a digitized society in which daily services are intertwined with real-name authentication~\cite{kim2018china}, yet NKDs must navigate these systems without safe legal identity. Many arrive with surveillance experience shaped by interpersonal monitoring and collective punishment rather than commercial platforms or metadata~\cite{warf2015hermit,lankov2014north}. Documented malware campaigns targeting this population confirm digital targeting is not hypothetical~\cite{klingner2021north}, producing a compound threat context in which origin-state pressure, host-country identity infrastructure, and constrained digital experience intersect.

Prior research examines NKDs' digital adaptation after resettlement in South Korea~\cite{noh2024starting,noh2025bridging}. We instead examine digital security before resettlement, when NKDs in China remain undocumented and often depend on intermediaries for digital access.

Guided by these gaps, we ask:
\begin{enumerate}[leftmargin=*, label=RQ\arabic*:]
  \item What digital threats do undocumented NKDs perceive and encounter in everyday practice?
  \item How do undocumented NKDs interpret and attribute digital threats when multiple actors, infrastructures, and dependency relationships overlap?
  \item What coping practices do they adopt, what protective scope do these practices provide under different threat models, and how do these practices change over time?
\end{enumerate}

We conducted semi-structured in-depth interviews with 19 undocumented NKDs in Northeast China, recruited over six months through trusted referral networks. Public recruitment could have exposed participants to detection by Chinese authorities or DPRK-linked actors, while unsolicited outreach could have been mistaken for spearphishing and deterred participation. Interviews were conducted in Korean and analyzed by a three-person research team using reflexive thematic analysis. Safety and data minimization were treated as central design constraints~\cite{bellini2024sok}.

Our findings identify three interacting conditions: transnational DPRK surveillance pressure; identity-linked infrastructure and intermediary dependency in China; and threat reasoning shaped by prior surveillance schemas and uneven familiarity with digital systems. These conditions compound one another. For example, lack of recognized legal identity can force reliance on employers or brokers for SIM cards, devices, and accounts, creating additional avenues for monitoring and control. Habituation follows two trajectories: \emph{adaptive habituation}, where learning enables more precise protection, and \emph{erosive habituation}, where the absence of visible harm weakens precautions without improving threat understanding.

We make four contributions: (1) an empirical account of how undocumented NKDs still residing in China attribute digital threats and reason about the protective scope of everyday practices; (2) an account of how overlapping actors and infrastructures complicate threat attribution; (3) evidence that prior surveillance schemas and feedback-poor environments influence threat reasoning and habituation; and (4) implications for security guidance, NGO support, platform design, and safer research with undocumented at-risk users. In such contexts, the conditions that make ordinary security advice actionable may themselves be absent.

\section{Background and Related Work}

This section situates our study within four bodies of literature. We first characterize undocumented NKDs in China as a compound threat context (\S\ref{2.1}), then review situated security practices among at-risk and displaced users (\S\ref{2.2}), examine work on mental models, habituation, and security advice (\S\ref{2.3}), and discuss methodological considerations for extreme at-risk populations (\S\ref{2.4}).

\subsection{North Korean Defectors in China as a Compound Threat Context}
\label{2.1}

Participants' digital safety problems are produced by the intersection of legal precarity, identity-linked infrastructure, transnational DPRK pressure, broker and employer dependence, and limited access to trustworthy security knowledge, making NKDs in China a compound threat context rather than a conventional digital-inclusion population.

\subsubsection{Legal Precarity, Digital Transition, and Identity Infrastructure}

Economic crisis in North Korea during the 1990s prompted large-scale border crossings into China~\cite{gunduz2017effect,lankov2004north,chang2008migration}. The Chinese government has largely treated NKDs as illegal economic migrants, leaving most without formal residency status and exposed to detention and forced repatriation~\cite{beck2007perilous,cho2013systemizing}. Most NKDs arrive from a society with state-controlled digital connectivity and little exposure to the open Internet or commercial platforms~\cite{warf2015hermit}, yet encounter in China a digitized environment in which communication, payment, and daily services are mediated through identity-linked services~\cite{castells2009mobile,kim2018china}. For undocumented NKDs, these systems become security-sensitive because participation requires identity documents they cannot safely present, and Chinese rules require network-log retention and real-name phone registration.\footnote{Cybersecurity Law of the People's Republic of China, art.~21; Provisions on the Registration of Real Identity Information of Telephone Users, arts.~3,~6,~11 (Ministry of Industry and Information Technology Order No.~25, 2013).} Phones, SIM cards, and account-recovery channels are therefore points at which access, exposure, dependency, and control converge. Unlike refugees studied in prior work, who adapt to unfamiliar digital systems while retaining some legal identity~\cite{simko2018computer,tran2024security,arunasalam2024understanding}, our participants must use identity-linked systems while lacking any safe legal identity through which those systems can be accessed.

\subsubsection{Transnational Repression and Digital Targeting}

Authoritarian states extend coercion beyond territorial borders through surveillance, proxy punishment of family members, coerced return, and cooperation with host-country institutions~\cite{moss2016transnational,moss2022going,glasius2018extraterritorial,tsourapas2021global}. NKDs in China are not protected from DPRK pressure by crossing a border; family members become leverage points and cross-border ties become channels of coercion. Civil-society spyware research and McAfee's RedDawn report confirm that politically motivated digital targeting of exile communities is not hypothetical~\cite{marczak2022hooking}.\footnote{McAfee RedDawn report on Sun Team malware targeting North Korean defectors through Google Play apps, including a food-information app.} Our study bridges a gap between political science accounts of cross-border coercion and technical malware reports by examining how these threats are experienced and reasoned about in everyday digital practice.

\subsubsection{Prior Research on NKDs' Digital Experiences}

Recent research has examined NKDs' digital transition after resettlement in South Korea, showing how defectors rebuild agency while encountering stigma and social comparison~\cite{noh2024starting,noh2025bridging,noh2026regaining}. Most directly, the 2025 NKDB report \emph{Caught in the Net} draws on a survey of 102 North Korean escapees who had lived in China, 20 in-depth interviews, and documentary analysis to characterize surveillance technologies, mobility restrictions, and human-rights consequences~\footnote{\url{https://en.nkdb.org/researchreport/?bmode=view&idx=168627987}}. Our study complements this institutional and infrastructural mapping by examining threat attribution, the conditional protective scope of everyday practices, and retrospectively reported adaptation among undocumented NKDs still residing in China. Our study examines a materially different phase: hidden residence in China, where using a phone may require another person's identity, contacting family may endanger DPRK relatives, and seeking help may require disclosing the very information participants are trying to protect. The gap is a threat-model gap between digital adaptation after legal recognition and digital safety before it.

\subsection{Situated Security Practices Among At-Risk and Displaced Users}
\label{2.2}

\subsubsection{Security Practices Under Social and Structural Constraint}

Security mechanisms fail when they do not fit users' goals, constraints, mental models, and social environments~\cite{whitten1999johnny,adams1999users}, and risk is further shaped by identity, location, relationships, and institutional position~\cite{warford2022sok}. At-risk users balance security against livelihood, mobility, and immediate safety~\cite{simko2018computer,mcgregor2017weakest,tran2024security, daffalla2021defensive, slupska2022they}. Across these studies, at least one stabilizing condition is typically present: a route to legal recognition~\cite{simko2018computer,tran2024security}, organizational resources~\cite{mcgregor2017weakest}, or an adversary distinguishable from the platform infrastructure~\cite{freed2017digital,matthews2017stories,tseng2022care}. We examine a population for whom several of these conditions may be absent at once, where removing one exposure channel can create another and technically correct advice may be structurally unactionable.

Neither changing security practices through experience nor declining vigilance under repeated exposure is unique to this population. Prior research with refugees and undocumented migrants shows that immediate needs and perceived benefits can push security and privacy concerns into the background, particularly when digital risks are poorly understood or recommended protections are difficult to enact~\cite{simko2018computer,tran2024security,guberek2018keeping}; habituation research likewise documents declining attention under repeated security warnings~\cite{vance2018tuning}. Our findings identify a different feedback condition within this broader pattern: without explicit warnings or reliable indicators of harm, participants sometimes treated the passage of time without visible consequences as evidence of safety. We distinguish this erosion of vigilance from knowledge-based recalibration in \S\ref{2.3}.

\subsubsection{Device Control, Third-Party Dependency, and Infrastructural Constraint}

IPV research shows that device and account control can become mechanisms of coercion~\cite{freed2017digital,matthews2017stories,leitao2019anticipating,tseng2022care,slupska2021threat}. In prior IPV work, the controlling actor remains analytically separable from the infrastructure provider. Our setting differs because the intermediary who enables digital access may also retain control over the device, account, or registered identity on which that access depends, collapsing access provision, surveillance capacity, and survival dependency into a single relationship.

\subsubsection{Digital Non-Use and the Cost of Disconnection}
HCI research frames digital non-use as a reflective or self-regulatory practice~\cite{baumer2013limiting,schoenebeck2014giving,satchell2009beyond}. In our setting, digital abstinence is a response to surveillance and legal exposure, yet the same channels that create exposure also carry aid information, community warnings, and security knowledge. When exposure channels and support channels are structurally co-located, non-use becomes a double-edged security practice rather than a straightforward reduction of attack surface.

\subsection{Mental Models, Habituation, and Security Advice}
\label{2.3}

\subsubsection{Security Mental Models and Schema Transfer}

Users reason about threats through partial and sometimes technically inaccurate frameworks~\cite{wash2010folk,wash2015too,krombholz2019if, hasegawa2024weird}. Our setting reveals a more specific mechanism: participants began not with incomplete digital models but with a coherent pre-digital surveillance schema formed under DPRK social surveillance. This schema made content-based risks cognitively available (sensitive speech, identifiable photos, device seizure) while rendering digital-specific traces invisible (metadata, server-side retention, device fingerprinting, possible account
linkage, and probabilistic social-graph inference). Drawing on schema theory and analogical transfer research~\cite{bartlett1932remembering,gick1983schema}, we analyze this as schema transfer producing structured visibility and invisibility rather than random ignorance.

\subsubsection{Habituation Beyond Warning Fatigue}

Prior habituation research shows that repeated warnings reduce attention over time~\cite{vance2018tuning} and that polymorphic warnings preserve attention through novelty~\cite{anderson2015polymorphic,felt2012android, vance2017we}. Our study extends this to a setting with no explicit warning prompts, where the relevant feedback signal is the presence or absence of visible harm. This produces two divergent trajectories: \emph{adaptive habituation}, where reduced fear follows concrete knowledge updates; and \emph{erosive habituation}, where reduced fear follows only the prolonged absence of adverse events. Under ambient surveillance, habituation bifurcates depending on whether users receive credible knowledge updates that allow risk recalibration.

\subsubsection{Security Advice, Compliance, and Structural Inefficacy}

Security advice fails when too costly, numerous, or mismatched to user goals~\cite{herley2009so,ion2015no,redmiles2020comprehensive,hielscher2023employees}. Our setting adds a structural limit: advice fails when users lack control over the device, account, registered identity, recovery channel, or help-seeking pathway needed to act on it, shifting the analysis from user compliance to the structural conditions under which advice becomes actionable.

\subsection{Research with Extreme At-Risk Populations}
\label{2.4}

Security research has increasingly recognized at-risk users as a central concern~\cite{warford2022sok}, with Bellini et al. synthesizing safer practices including threat modeling the research process, minimizing data collection, and carefully scoping publication~\cite{bellini2024sok}. Following case-study methodology, extreme cases make mechanisms visible that moderate contexts obscure~\cite{flyvbjerg2004five}. Undocumented NKDs in China stress-test core usable security assumptions: that users can identify adversaries, control devices, obtain trustworthy advice, observe feedback, and act on recommendations. The failure of these assumptions under extreme conditions clarifies the boundary conditions of existing theory.

\section{Method}

\subsection{Research Design}

We conducted a qualitative interview study with undocumented NKDs in Northeast China to examine how participants understood digital threats, developed protective practices, and adapted phone and platform use under legal precarity, transnational repression, and infrastructural dependency. The three-person research team included a native Korean speaker who led participant communication, consent, interviewing, transcription, and translation. Language access, cultural interpretation, and trust-sensitive communication were central design requirements.

We chose semi-structured in-depth interviews to understand participants' lived security reasoning rather than measure predefined behaviors. Surveys were poorly suited to a hidden, legally vulnerable population whose practices and threat interpretations were not well characterized, and risked broader recruitment and identifiable response traces. We avoided experiments, device inspections, and forensic audits because intrusive access to phones and accounts could create risks if discovered by employers, brokers, or authorities. Semi-structured interviews provided comparable coverage while allowing flexibility to follow individual experiences and avoid unsafe questioning.

We focused on undocumented NKDs because legal precarity, identity-linked infrastructure, employer dependence, and DPRK transnational pressure intersect, providing an extreme case for examining security practices without assumed legal recognition, independent device control, or trusted advice channels. Rather than seek statistical generalization, we identify mechanisms of security reasoning within this high-risk context.

Interviews lasted 60–85 minutes and covered migration background, everyday phone use, perceived threats, technical understandings, coping practices, and change over time. Modality followed participant safety and preference: interviews were conducted in person only when participants considered the setting safe, otherwise by telephone or Tencent Meeting. We did not require unfamiliar security tools because they could create suspicion or new traces.

\subsection{Participants}

\subsubsection{Recruitment Process}

We recruited participants entirely through trusted personal networks via snowball sampling in Northeast China. Conventional recruitment channels such as posted flyers, online announcements, or institutional calls would have posed unacceptable safety risks by exposing participants' identities, locations, or association with defector networks. We initiated contact through a small number of trusted intermediaries with pre-existing relationships in local defector communities. Intermediaries first confirmed preliminary willingness before the research team proceeded with screening and scheduling. Each recruited participant could refer additional individuals who met the inclusion criteria and voluntarily consented to be approached. All contact, screening, and scheduling used communication channels the intermediary and prospective participant already considered safe. No personal identifying information was collected or transmitted before participants explicitly agreed to proceed. Recruitment spanned six months across multiple cities and counties in Northeast China.

\subsubsection{Inclusion Criteria}

We included participants meeting all of the following criteria at the time of recruitment:

\begin{enumerate}[leftmargin=*, label=(\arabic*)]
  \item \textbf{Origin}: Born in and having resided in the DPRK prior to departure.
  \item \textbf{Current residence}: Living in China without formal legal residency status recognized by Chinese authorities.
  \item \textbf{Smartphone use}: Having used a smartphone or mobile internet service at any point during their time in China.
  \item \textbf{Age}: Aged 18 years or older.
  \item \textbf{Voluntary participation}: Willing and able to provide informed consent without coercion.
\end{enumerate}

We excluded individuals who had obtained formal residency or citizenship in a third country, or who were in a UNHCR-managed protection process, as their threat environment differed materially from that of undocumented defectors in China.

\subsubsection{Recruitment Outcomes}

During the six-month recruitment period, 19 eligible participants were safely reached, consented, and completed interviews. The sample was not designed to be statistically representative; it reflects the ethical and practical constraints of qualitative research with a hidden and legally vulnerable population. The sample included 12 women and 7 men, with ages ranging from the early twenties to the late fifties and the majority between 25 and 40. Educational attainment was predominantly at the middle-school level (11 participants), with 2 holding primary-school credentials, 2 holding high-school credentials, and 4 declining to disclose. Time in China ranged from under one year to 12 years, with a median of 2.5 years. This temporal heterogeneity supported comparison across participants at different stages of adaptation, though we do not claim longitudinal inference from cross-sectional data. To protect participants in accordance with IRB requirements, specific personal information is not disclosed.

\subsubsection{Ethical Considerations}

This study involved a highly vulnerable population whose participation entailed material legal and physical risk. We treated ethical protection as a central design constraint rather than a procedural formality, operationalized across five dimensions.

\paragraph{\textbf{Informed Consent}}
We obtained voluntary informed consent before data collection. Consent was conducted orally in Korean given varying literacy levels, with written records created only when participants explicitly agreed. We explained the study's purpose, data handling, the right to withdraw without consequence, and identity-protection measures. Each participant received 400 RMB (approximately 57 USD), not tied to answering any particular question. Given the lengthy interviews, the sensitive nature of the research, and the significant burden on participants, the compensation exceeds local standard rates.

\paragraph{\textbf{Anonymization and Data Minimization}}
We did not collect real names, precise locations, employer details, route information, or family information beyond what was necessary to understand participants' accounts. Each participant was assigned a pseudonymous identifier (P1 through P19) at first contact, and we report age ranges rather than exact ages throughout, generalizing or omitting any details that could enable re-identification. For quotations where participant identifiers could enable contextual inference or involve particularly sensitive information, the identifiers were further anonymized (Numbered Anonymization) 

\paragraph{\textbf{Risk Assessment and Harm Minimization}}
Before each interview, the interviewer confirmed the participant was in a safe location with no uninvolved third parties present, and avoided locations identified as monitored or where coercive actors were nearby. Participants were reminded throughout that they could skip or stop at any time; reluctance to engage with any topic was not pursued. A distress response protocol, including a prepared list of support contacts, was maintained throughout.

\paragraph{\textbf{Data Retention and Disposal}}
Recordings were transferred immediately after each interview to air-gapped devices accessible only to core team members, then deleted from the recording device. Transcripts were stored under participant codes and de-identified after active analysis. Raw transcripts, recordings, and operational details are not shared with third parties in any form that could enable re-identification.

\paragraph{\textbf{Institutional Review}}
This study has been reviewed and approved by the authors' institutional review ethics committee, with the review covering study design, participant recruitment, informed consent, data processing, and risk mitigation measures.

\subsection{Data Collection}

Before each interview, we contacted participants through intermediary-approved channels to agree on time, location, and communication method, and explained the study's purpose and withdrawal rights at this stage so participants could decide without pressure. Three interviews were conducted in person (P1, P12, P14); all others participated by telephone or Tencent Meeting, with modality selected based on participants' safety assessment and preference. In the case of P12, a risk factor identified during the first scheduled session led us to defer the interview.

The Korean-fluent researcher conducted all interviews using a semi-structured protocol across seven thematic modules, adjusting probe order and depth in response to each participant's answers. Question sequencing followed a principle of progressive depth, opening with basic demographic questions before moving to sensitive topics. Modules covered North Korean surveillance experience and baseline cognition; living conditions and privacy concerns in China; device-level practices including SIM acquisition, real-name registration strategies, data deletion, and location tracking awareness; and participants' subjective threat models. Structured knowledge probes assessed understanding of specific technical mechanisms, including whether disabling Wi-Fi prevents location tracking, whether deleted photos remain recoverable, whether translation apps transmit entered text, and what protections encrypted tools such as Telegram do and do not provide. A final module examined how risk awareness and phone-use habits had changed since arriving in China. Each interview closed with a decompression segment in which the interviewer invited final remarks and provided information about available community support resources.

\subsection{Data Analysis}

We analyzed transcripts using thematic analysis, aiming to develop explanatory themes about participants' threat reasoning, protective practices, and adaptation over time rather than to quantify behavior prevalence. Three researchers participated across five stages.

First, all three researchers read all translated transcripts in full, with the Korean-fluent researcher flagging uncertain passages and culturally specific expressions for team discussion. Each researcher wrote preliminary analytic memos to document first impressions and guard against preconceptions. Second, each researcher independently coded transcripts at the level of meaning units using descriptive codes close to participants' accounts, covering behaviors, perceived threat sources, technical explanations, uncertainties, and expectations, evaluations of protective strategies, and accounts of change over time. Multiple codes could be assigned to the same segment. Third, the team compared codes through analytic meetings, treating disagreement as an analytic resource and returning to source wording to revise definitions, merge or separate codes, and document unresolvable ambiguity. Fourth, codes were iteratively aggregated into higher-level categories, with the team moving between excerpts and candidate categories to verify coverage and resolve overlaps. Fifth, themes were constructed by identifying the analytic claim connecting each category cluster, checked against negative and discrepant cases. Five recurrent patterns anchored the final themes: mixed attribution of surveillance sources; transfer of non-digital surveillance schemas into digital contexts; differences between participants' expected protection and our threat-relative assessment of protective scope; infrastructural dependency on employers, brokers, and identity-bound systems; and temporal relaxation of risk perception with length of residence.

By the fourteenth participant, newly emerging categories had decreased substantially and themes had stabilized. We did not treat saturation as the decisive criterion, however, given the rarity and vulnerability of this population and the safety constraints on recruitment. All subsequently recruited participants (P15 through P19) were retained to test framework robustness and capture variation among longer-term residents. We treat the sample as providing analytic sufficiency for mechanism-building, not statistical representativeness.

Throughout analysis, we engaged in sustained reflexive practice, discussing how linguistic positions, cultural backgrounds, and prior assumptions might shape interpretation. The Korean-fluent researcher contributed source-language and cultural interpretation while the broader team challenged emerging claims from security, privacy, and HCI perspectives, helping us analyze participants' practices as shaped by legal vulnerability and infrastructural dependency rather than as mere technical errors.

Our assessment of coping practices was mechanism-level rather than forensic. We did not inspect participants' devices or independently validate specific surveillance incidents. Prior work shows that a security practice's perceived effectiveness and its actionability for a given user are empirically distinct and often diverge~\cite{redmiles2020comprehensive}. We similarly assessed each reported practice along two separate dimensions: the technical mechanism it could plausibly address, and whether participants could sustain or control the practice given their infrastructural and legal constraints. Specifically, for each reported practice we identified the mechanism it was intended to address (e.g., metadata retention, device or account linkage, cellular-network location signals) and characterized its plausible protective scope by cross-referencing established findings on that mechanism~\cite{narayanan2009anonymizing, mayer2016evaluating, tomasin2021location}. We therefore distinguish participant-reported experience, the technical mechanism a practice targets, and our own interpretation of what a practice can and cannot protect against, rather than issuing a single effective/ineffective judgment.

\section{Results}
\label{sec:findings}

Participants' security experiences reflected overlapping surveillance pressure, infrastructural dependency, and uncertainty in threat reasoning. Their coping practices varied in protective scope, while habituation followed two trajectories: a minority developed more precise protective behaviors through knowledge updates, whereas a larger group gradually relaxed early precautions because ``nothing has gone wrong so far,'' falling into \emph{erosive habituation}.

\subsection{Interacting Conditions Shaping Digital Security}
\label{subsec:interacting-conditions}

We organize the findings around three empirically distinct but interacting conditions: transnational surveillance pressure from the DPRK, identity-linked infrastructure and intermediary dependency in China, and threat reasoning under digital and adversarial uncertainty. They overlap through shared infrastructures and dependency relationships, jointly structuring exposure and participants' interpretations of risk.

\subsubsection{Transnational Surveillance Pressure from the DPRK}
\label{subsubsec:dprk-pressure}

North Korean surveillance pressure operated through participants' internalized experience of social surveillance before departure and ongoing digital-tracking threats after defection.

\paragraph{\textbf{(1) Internalized Surveillance Consciousness}}
Most participants' security and privacy awareness did not form in China. The \emph{inminban} (people's or neighborhood unit) system and the distinct five-household surveillance system embed monitoring, reporting, and ideological control within everyday neighborhood life~\footnote{\url{https://www.kinu.or.kr/eng/module/report/view.do?idx=128276&nav_code=eng1674806000}}. These institutions made surveillance a familiar baseline before participants left North Korea and shaped the schemas through which some later interpreted digital environments.

\begin{displayquote}
\textit{``Nothing is safe. And I don't just mean digital systems; I mean everything. You have to consider what impact a piece of information might have, how it might be interpreted by people. I think this is something we defectors share: you have to be suspicious of everything.''} (P4)
\end{displayquote}

\paragraph{\textbf{(2) Communications Surveillance}}
The North Korean State Security Department (\emph{Bowibu}, bowibu) monitors call records and communication content of defectors on the run, focusing on contacts with numbers in North Korea and South Korea. Nine participants mentioned this pressure. North Korean security agencies also use family members remaining in the DPRK as leverage; some participants recognized that relatives who had not yet been prosecuted may already have become coerced informants.

\begin{displayquote}
\textit{``Making calls carries risk, especially to Korean or North Korean numbers. Right after I escaped, my aunt called me and spoke in a very strange way, said things that couldn't possibly have been her own words. I knew she had been taken over.''} (P14)
\end{displayquote}

\paragraph{\textbf{(3) Mobile Spyware Targeting the Defector Community}}
North Korea disguises malicious programs as practical utilities such as recipe, health, and safety apps and distributes them through community channels, silently collecting SMS messages, contacts, call logs, GPS coordinates, and photos. These operations exploit defector communities' reliance on peer networks to relay information.

\begin{displayquote}
\textit{``Some Korean-language apps, and apps recommended by other defectors, are quite possibly problematic. They likely record information such as who you messaged, your location, the time, photos, and so on.''} (P1)
\end{displayquote}

\paragraph{\textbf{(4) Social Media Identity Recognition}}
Most participants mentioned that North Korea uses facial-recognition tracking on social media platforms and extends outward from known defectors' accounts through social graphs to identify fugitives whose identities are unknown. After certain platforms began mandatorily displaying IP addresses, some participants stopped using those platforms entirely.

\begin{displayquote}
\textit{``I don't post photos; not a single one. I'm afraid of Pyongyang. I've heard they have a way to find people through photos; I don't know exactly how, but I don't want to take that risk. Later, when some social media started to force-display IP addresses, I stopped using some of them, and on the ones I still use I never speak.''} (P11)
\end{displayquote}

\paragraph{\textbf{(5) Device Seizure and Data Extraction}}
Four participants noted that upon arrest, call records, chat logs, contact lists, and photos stored on a device could constitute an evidentiary chain implicating other defectors in China, creating sustained pressure on their digital behavior.

\begin{displayquote}
\textit{``I've developed the habit of regularly deleting chat records. If I'm ever caught, I can't let them use my phone to find other people. I can easily imagine what they would do with my phone and my data if I were arrested.''} (P9)
\end{displayquote}

\subsubsection{Identity-Linked Infrastructure and Intermediary Dependency in China}
\label{subsubsec:identity-dependency}

China's digital infrastructure created four pressures in participants' everyday lives: exclusion caused by absent identity documents, platform retention of communications behavior, facial-recognition exposure in public spaces, and digital control through employers and brokers. A minority with explicit plans to leave China (e.g., P10) expressed these pressures more intensely.

\paragraph{\textbf{(1) Real-Name Registration: No Identity Means Digital Exclusion}}
For participants lacking legal identity, China's real-name registration system pushed them to the digital margins. SIM cards, WeChat, Alipay, and other mainstream services require binding a valid identity document, preventing participants from accessing them under their own names.

\begin{displayquote}
\textit{``In principle, here, without a real identity, there are almost no digital devices one can fully use. Everything, including internet services, phone numbers, and countless apps, requires an identity to be bound. Without an identity, or when you don't want to expose yours, you can only use very limited functions.''} (P16)
\end{displayquote}

\paragraph{\textbf{(2) Platform Surveillance: Comprehensive Retention of Communications Behavior}}
Participants described platforms as retaining communication content, device IMEI numbers, usage-time patterns, and geographic location trajectories, creating behavioral records beyond message content itself. Who they contact and when can therefore also constitute analyzable information.

\begin{displayquote}
\textit{``Whatever app, whatever platform you use, there are traces. Someone will know. Once I accidentally said the wrong thing and got targeted; fortunately that account was completely erased and I moved somewhere else to live.''} (P10)
\end{displayquote}

\paragraph{\textbf{(3) Facial Recognition: Pervasive Identity Verification}}
China's large-scale deployment of facial-recognition systems at public transit nodes, commercial districts, and residential community entrances made everyday movement a potential source of identity exposure. Participants could neither avoid public space nor control whether their facial features were matched against existing records.

\begin{displayquote}
\textit{``There are more and more places I don't dare go. Train stations, subway entrances, those busy malls: I try to avoid them. Not because I've done anything, but because I don't know whether my face is already in that system. I've gone back to living in the countryside now; it's a bit better, with less surveillance.''} (P5)
\end{displayquote}

Pandemic-era prevention-and-control measures concentrated this pressure. Six participants described the tension, between 2019 and 2022, between forced identity verification and their lack of identity documents, during which digital and physical control overlapped.

\begin{displayquote}
\textit{``During those three years, I basically couldn't take a step. Couldn't do anything, couldn't use anything. Just staying indoors, in a cellar, even hiding in the hillside woods out back.''} (P19)
\end{displayquote}

\paragraph{\textbf{(4) Employer and Broker Relationships: Protection as Control}}
Most participants' phones and SIM cards were purchased and registered in an employer's or broker's name. Some employers used this access for direct control by inspecting phone contents or confiscating devices. The relationship participants relied on to evade state surveillance could therefore itself become another layer of surveillance.

\begin{displayquote}
\textit{``He has the right to look, because that number is registered in his name and that phone was bought with his money. He went through my phone and I couldn't say anything; I didn't even dare ask what he'd seen. The feeling is terrible, but what can you do? You live at his place; you depend on him.''} (P13)
\end{displayquote}

\subsubsection{Threat Reasoning Under Digital and Adversarial Uncertainty}
\label{subsubsec:threat-reasoning}

Participants' threat reasoning reflected four related processes: limited prior exposure to smartphones and the open internet, transfer of interpersonal surveillance schemas into digital contexts, uncertainty about which actors can access information, and generalization of risk under persistent ambiguity.

\paragraph{\textbf{(1) Pre-Digital Background and Cognitive Starting Point}}
The vast majority of participants had almost no experience with smartphones or the open internet before leaving North Korea; only a small number had accessed smuggled foreign media through informal channels. They therefore encountered smartphones with threat schemas built on non-digital surveillance experience rather than prior knowledge of digital systems.

\begin{displayquote}
\textit{``Ordinary North Koreans have almost no experience using smartphones or the internet. Personally, the first time I came into contact with a phone was in Dandong. Of course there are exceptions, such as Korean films.''} (Numbered Anonymization)
\end{displayquote}

These schemas made some risks highly salient while providing fewer conceptual tools for reasoning about less visible mechanisms such as metadata retention, device linkage, and server-side persistence.

\paragraph{\textbf{(2) Migration of Threat Schemas: Over-Analogy and Schema Gaps}}
Participants understood digital devices through the interpersonal-surveillance logic of ``spoken words can be overheard,'' obscuring risks such as metadata retention and device fingerprint linkage that lack interpersonal counterparts. Focusing on communication content left automatically retained metadata, including call timing, duration, and counterpart numbers, unaddressed.

\begin{displayquote}
\textit{``I thought as long as I didn't say anything sensitive it would be fine. I only gradually learned that who you contact and how often you contact them is itself information. I discovered this slowly over time; fortunately the leakage of that information didn't get me sent back.''} (Numbered Anonymization)
\end{displayquote}

For risks with no counterpart in prior surveillance experience, most participants lacked not only coping strategies but also a cognitive framework for perceiving them. GPS coordinates embedded in photos and behavioral data collected in the background by apps could therefore remain cognitively invisible and absent from subjective risk assessment.

\begin{displayquote}
\textit{``A friend warned me and showed me a photo I had sent, saying it contained my location at the time I took it. I said that's impossible; a photo is just an image. After they showed me, I was speechless. I had lived in that place for two years and sent lots of photos.''} (P12)
\end{displayquote}

\paragraph{\textbf{(3) Threat-Attribution Uncertainty}}
Participants often lacked enough information to reliably attribute a particular exposure or surveillance capability to one actor. Chinese state surveillance, cross-border tracking by North Korean security services, employer control, platform visibility, and community leakage could therefore appear as a diffuse condition of being watched. This uncertainty was produced by overlapping actors, infrastructures, and dependencies whose access boundaries were difficult for participants to observe, rather than simply by a mistaken belief about one adversary.

\begin{displayquote}
\textit{``I just feel someone is watching me; I can't say exactly who. Chinese police, the North Korean side, my connections: all are possible. So I'm careful about everything, but I can't really say what specifically I'm being careful about.''} (Numbered Anonymization)
\end{displayquote}

\paragraph{\textbf{(4) Threat Generalization Under Persistent Uncertainty}}

Participants treated all digital behavior as a potential threat and could not distinguish high-risk operations from routine use. This reinforced broad, non-actor-specific threat reasoning: the more diffuse the perceived threat source, the wider the precaution and the harder it became to accumulate experience about specific threat mechanisms.

\begin{displayquote}
\textit{``When I first arrived, I didn't even dare accept WeChat friend requests. I didn't know who it was, didn't know their purpose. I just felt unsafe. That feeling didn't go away for a very long time.''} (P15)
\end{displayquote}

\subsection{Participants' Coping Strategies}
\label{subsec:coping}

Participants adopted coping practices targeting different perceived harms, including communication exposure, identity linkage, device inspection, location visibility, and uncertainty itself. Because the same practice can protect against one threat while doing little against another, we assess these practices at the level of specific threat mechanisms rather than classifying them globally as effective or ineffective. Our assessment compares participant-reported practices with the technical mechanisms they were intended to address and with established security and privacy research. Over time, habituation and knowledge networks further shaped how these practices evolved.

\begin{table*}[t]
\centering
\caption{Threat sources, accessible information or capabilities, participant concerns, observed coping practices, and our mechanism-level assessment of protective scope. Prevalence counts reflect participants who described the threat or practice ($N$=19). Assessments characterize the scope and limitations of reported practices rather than globally classifying them as effective or ineffective.}
\label{tab:overview}
\setlength{\tabcolsep}{3.5pt}%
\resizebox{\textwidth}{!}{%
\begin{tabular}{@{}p{2.4cm}p{4.1cm}p{4.2cm}p{4.4cm}p{5.8cm}@{}}
\toprule
\textbf{Threat actor / source} & \textbf{Capability or accessible information} & \textbf{Participant concern} & \textbf{Observed coping practice} & \textbf{Protective scope and limitation} \\
\midrule
DPRK-linked surveillance & Communications monitoring; family-based leverage & Contacts or sensitive content may expose participants or relatives & Call relaying, code words, minimal duration & May reduce direct content disclosure or local linkage; carrier-side metadata can remain available \\
DPRK-linked surveillance & Community-distributed mobile spyware collecting device data & Apps may expose messages, contacts, location, or photos & Official-app-store-only installs & Reduces exposure to sideloaded malware but does not address dual-use apps that pass official app-store review and evade conventional anti-spyware detection~\cite{chatterjee2018spyware} \\
DPRK-linked surveillance & Social-media identity and graph analysis & Photos, accounts, or connections may reveal identity & No facial photos; platform exit after IP display & Reduces some directly disclosed identity signals; body pose, clothing, and photo context can still support high-accuracy re-identification even with faces fully obscured~\cite{oh2016faceless} \\
DPRK / authorities after device seizure & Local device records and contacts & Stored data may implicate the participant or others & Routine local deletion & Reduces information visible through casual inspection, but forensic extraction can still recover residual artifacts (cached files, timestamps, directory structures) from the device itself even after deletion~\cite{grispos2015recovering} \\
\midrule
PRC identity infrastructure & Real-name registration required for SIMs and mainstream services & Lack of recognized identity restricts independent digital access & Employer/broker-registered SIMs and devices & Enables access but creates dependency on the registering intermediary \\
Platforms / carriers & Account, device, behavioral, and location traces & Persistent traceability across accounts and use patterns & Periodic account cycling; re-adding contacts on the same device ($n=11$) & Can separate some account-level history; protection is limited when device or social-graph continuity remains~\cite{narayanan2009anonymizing} \\
PRC public-space infrastructure & Facial recognition and identity-linked mobility controls & Everyday movement may produce identity exposure & Route planning; rural relocation & May reduce encounters with dense monitoring environments but cannot eliminate public-space exposure, as China's camera-network expansion has extended to both urban and rural areas~\cite{ma2026keeping} \\
Employers / brokers & Registered ownership, device access, inspection, and confiscation & Monitoring and control through a survival dependency & Often no feasible active response & Dependency constrains the participant's ability to refuse, exit, or independently reconfigure the device/account relationship \\
\midrule
Digital trace mechanisms & Metadata retained independently of communication content & Participants may focus on content while underestimating relational or behavioral traces & Selective content avoidance & Reduces content disclosure but does not address metadata-based inference~\cite{mayer2016evaluating} \\
Network / storage mechanisms & Cellular location signals; server-side retention & Participants may expect Wi-Fi disabling or local deletion to remove broader traces & Disabling Wi-Fi; clearing local data & Can reduce some Wi-Fi interactions or locally visible records; does not eliminate cellular location signals or server-side copies~\cite{tomasin2021location} \\
Diffuse / unattributed threat & Uncertain boundaries among actors and infrastructures & Generalized concern that digital activity itself creates exposure & Digital abstinence & Reduces some digital trace production but also cuts off aid, warnings, support networks, and security knowledge \\
\bottomrule
\end{tabular}%
}
\end{table*}

\subsubsection{Practices Addressing Communication, Identity, and Physical Exposure}
\label{subsubsec:concrete}

Participants' strategies in response to North Korean tracking threats and China's surveillance infrastructure broadly comprised communication protection, identity concealment, and physical avoidance.

\paragraph{\textbf{(1) Communication Protection}}
Participants generally adopted two measures: call-path obfuscation and content control. The former included routing calls through intermediaries and minimizing call duration; the latter involved avoiding references to identity, location, or sensitive relationships.

\begin{displayquote}
\textit{``When I make calls, I don't use my own number; I have someone else relay it. And I never mention specific places on the phone. I use code words, say `that place,' `that thing,' never real names.''} (P2)
\end{displayquote}

Faced with device seizure risk, periodically clearing local data was habitual among all participants, covering chat logs, call records, and photos. Some followed fixed schedules; others cleared data immediately after each call or meeting.

\begin{displayquote}
\textit{``Every time I finish a call I delete it; every time I meet someone I clear the chat records. It's a habit; something that has to be done.''} (P6)
\end{displayquote}

\paragraph{\textbf{(2) Identity Concealment}}
Faced with social-media identity-recognition threats, participants strictly controlled facial information, refusing to post identifiable photos and minimizing account information. Some stopped using platforms once they began mandatorily displaying IP addresses.

\begin{displayquote}
\textit{``My profile picture has never been me; in fact never a real photo. My username isn't my real name; I've filled in no personal information at all. When people ask for photos, I say I don't have any. I can't give them.''} (P7)
\end{displayquote}

Faced with mobile spyware threats, participants were wary of non-official-channel applications. Some strictly limited downloads to the Apple App Store and rejected software recommendations from strangers or circulated within the community.

\begin{displayquote}
\textit{``I never click on any software I'm not familiar with, especially anything outside the Apple App Store. Even if it's sent by someone I know, I won't install it. I don't know what's inside, and I don't trust it.''} (P3)
\end{displayquote}

\paragraph{\textbf{(3) Physical Avoidance}}
Some participants planned routes to avoid public transit nodes and commercial areas with dense facial-recognition coverage. This was especially prominent among urban residents; some ultimately returned to rural areas to reduce exposure.

\begin{displayquote}
\textit{``Before going out now I think about the route. Whichever road has fewer cameras, that's the one I take. Sometimes I have to walk much farther, but I think it's worth it.''} (Numbered Anonymization)
\end{displayquote}

Faced with control through employer and broker relationships, participants had few feasible coping strategies. They recognized that employers could access their digital activities but, given the dependencies sustaining their survival, could not readily refuse or constrain that power. This lack of feasible response is itself a finding: dependency can make a threat recognizable without making it actionable.

\begin{displayquote}
\textit{``I know he can look; I know he has looked. But what can I say? I live at his place; my number is registered through him. What can I do?''} (Numbered Anonymization)
\end{displayquote}

\subsubsection{Practices with Mismatched or Limited Protective Scope}
\label{subsubsec:transformed}

Several practices had clear protective intent but addressed only part of the associated threat. Rather than treating them as globally ineffective, we examine their plausible protective scope and the mechanisms left unaddressed.

\paragraph{\textbf{(1) Borrowing Devices to Make Calls}}
Participants generally believed that borrowing a phone for sensitive calls prevents records from appearing on their own device, thereby severing the link to their identity. This confuses device-local records with network-layer metadata. Carrier records exist independently of local storage and include calling and called numbers, duration, and timing. Borrowing therefore leaves network-layer traces and introduces the device owner as another party with knowledge of the interaction.

\begin{displayquote}
\textit{``Some calls are inconvenient to make on my own phone, so I borrow someone else's. That way the call record isn't on my phone; if they check me, they won't find it. Whether the carrier has a record? I hadn't thought about that.''} (Numbered Anonymization)
\end{displayquote}

\paragraph{\textbf{(2) Periodic Account Cycling}}
Some participants believed that periodically abandoning old accounts and registering new ones breaks accumulated digital-identity continuity. Account cycling can separate some account-level history, but its protective scope is limited when accounts remain linked through the same device, phone number, or social connections: overlapping social-graph structure alone can support re-identification even without usernames and other identifiers~\cite{narayanan2009anonymizing}. Eleven participants reported re-adding contacts and switching accounts on the same device, leaving continuity signals that can support linkage.

\begin{displayquote}
\textit{``I change my WeChat account roughly every three months. After switching I feel refreshed, like a fresh start. I do add my previous contacts back, gradually, one by one. I think this should still be useful, since the account is different after all.''} (Numbered Anonymization)
\end{displayquote}

\paragraph{\textbf{(3) Selective Content Avoidance}}
Most participants strictly avoided disclosing identity, location, and sensitive relationships in message content, but seven reported no practices aimed at limiting metadata exposure. This reduced content disclosure but did not address metadata-based inference: contact patterns, frequency, timing, and location trajectories could still reveal social relationships and identity without message content, and communication metadata alone can support highly accurate inferences about relationships, location, and sensitive personal traits~\cite{mayer2016evaluating}.

\begin{displayquote}
\textit{``On WeChat I never share real information; I don't say where I am, don't say who I am, don't mention any specific names or places. I think even if they look they can't find me. \ldots Can they really infer things just from who I contact and how often? I don't think that's very likely.''} (Numbered Anonymization)
\end{displayquote}

\paragraph{\textbf{(4) Disabling Wi-Fi and Clearing Data}}
Fourteen participants described disabling Wi-Fi to reduce location traceability and clearing local chat records to eliminate stored data. Disabling Wi-Fi can reduce Wi-Fi-based interactions but does not eliminate cellular-network location signals, which operators can obtain independently of user-facing applications or settings~\cite{tomasin2021location}. Clearing local records reduces information immediately visible through device inspection but leaves server-side copies and independently retained metadata outside users' control. Both practices therefore had narrower protective scope than participants often expected.

\begin{displayquote}
\textit{``I clear my chat records once a week and turn off Wi-Fi whenever I go somewhere sensitive. It's my habit; I feel at ease after doing it. \ldots Tencent still has a copy after I delete it? Cellular network still locates me after I turn off Wi-Fi? That can't be right; surely it's not that advanced \ldots''} (Numbered Anonymization)
\end{displayquote}

\paragraph{\textbf{(5) Digital Abstinence}}
The extreme form of threat generalization was digital abstinence: some participants avoided smartphones or reduced digital contact to an absolute minimum. Their logic was internally consistent: if digital devices enable tracking, rejecting them eliminates those exposure vectors. However, abstinence also cut participants off from humanitarian aid, legal aid, defector mutual-aid networks, and digital security knowledge. Several participants learned during interviews, for the first time, about a grassroots underground support organization that had operated in their city for years. Digital avoidance had prevented this information from reaching them.

\begin{displayquote}
\textit{``The less I use, the less they can trace. I've always thought that. I use a phone that can only make calls. \ldots You're saying there's an organization in this city that helps defectors? I've lived here for a while now and no one ever told me. I don't use WeChat, so I don't know anything.''} (P17)
\end{displayquote}

\subsubsection{Long-Term Dynamics: Habituation and Knowledge Networks}
\label{subsubsec:longitudinal}

Compared with participants with shorter residence periods (P1, P3 to P10, P17; all under three years), those with longer residence (P2, P11 to P16, P18, P19; all at least three years) generally exhibited a \emph{habituation response}: gradually weakening broad precaution and perceived threat. Beneath this pattern were two mechanistically distinct trajectories with different long-term consequences.

\textbf{Adaptive habituation} occurred among participants who actively updated their digital security knowledge through NGO training or diverse information channels (only P14 and P19). Reduced perceived risk could be linked to specific knowledge-update events and was accompanied by a shift from broad, undifferentiated avoidance toward precise, targeted protection.

\begin{displayquote}
\textit{``Not everything needs to be feared. I think two things are very important: experience and learning. In terms of experience, some people got caught because of certain behaviors, while my own behaviors haven't led to my being caught, so gradually I've developed a better understanding of what can and can't be done. Beyond that, I think even more important is proactive learning. As the proverb says: `Know how the devil moves, and you have the means to avoid it; you don't have to be afraid every moment.'\,''} (Numbered Anonymization)
\end{displayquote}

\textbf{Erosive habituation} was more common (P2, P12, P13, P15, P16, and P18). Participants in this group said they had gradually relaxed their initial precautions because they had not observed any adverse consequences, even though their understanding of surveillance mechanisms had not improved.

\begin{displayquote}
\textit{``For the first two years I posted nothing, added no one, said nothing. Then gradually I got used to it; nothing happened, so I thought maybe it's not that serious. Did some new understanding make me feel the risk was lower? No. I just got less scared as time went on.''} (Numbered Anonymization)
\end{displayquote}

Habituation was not pronounced among all longer-term residents. One participant remained highly vigilant toward uncertain threats. Although they lacked a precise model of how digital threats operated, prior encounters with state surveillance furnished knowledge that helped them recognize some threats. An assets-based lens foregrounds this situated knowledge rather than reducing the participant's account to low digital literacy~\cite{wongvillacres2020culture,wongvillacres2022elevating}.

\begin{displayquote}
\textit{``I suffered the most severe persecution in North Korea; it was a purely dark experience. I cherish my current life enormously, and I have to avoid every possible risk. That is why I have been working so hard throughout these years.''} (Numbered Anonymization)
\end{displayquote}

Even so, signs of habituation were present in this participant's account: \textit{``I'm really, really tired. I'm now thinking about finding some channel to properly learn which behaviors are too dangerous, so my burden can get a little lighter. Otherwise it's truly exhausting.''}

Similarly, among participants with fewer than three years of residence, some had already begun to show habituation responses.

\begin{displayquote}
\textit{``Gradually, I'm not as scared as I was at the beginning. Still quite scared, but you have to live your life.''} (P8)
\end{displayquote}

\section{Discussion}
\subsection{Undocumented NKDs as a Stress Test for At-Risk User Research}

Our findings show that undocumented NKDs in China face a compound threat context in which DPRK transnational pressure, China's identity-linked infrastructure, intermediary dependence, constrained digital knowledge, and community-mediated information channels interact. This reinforces prior work showing that at-risk users adapt security practices to resource and safety constraints~\cite{simko2018computer,mcgregor2017weakest,sambasivan2018privacy,tran2024security}, particularly under asymmetric control over devices and accounts~\cite{freed2017digital,matthews2017stories,tseng2022care,leitao2019anticipating}. Security advice can be difficult to act on under such constraints~\cite{herley2009so,ion2015no,redmiles2020comprehensive,hielscher2023employees}, while repeated exposure may reduce vigilance~\cite{vance2018tuning,anderson2015polymorphic,felt2012android}.

Beyond documenting an understudied population, this case tests the limits of common assumptions in usable-security research. Existing work has examined refugees, migrants, journalists, activists, IPV survivors, low-resource users, and other at-risk groups~\cite{warford2022sok,simko2018computer,mcgregor2017weakest,freed2017digital,tran2024security}. These studies have expanded security research beyond generic user models. However, many contexts retain at least one stabilizing condition, such as a path to legal recognition, some control over devices or accounts, organizational support, or a relatively trusted information channel.

Undocumented NKDs in China often lack these conditions simultaneously. Without recognized legal identity, digital participation can entail identity exposure, while access to phones, SIM cards, accounts, and support may depend on employers, brokers, or other intermediaries. Community information channels may likewise provide warnings while carrying rumors, coercion, or malicious software. Consequently, users may be unable to identify a primary adversary, control relevant devices or accounts, safely seek advice, or implement technically sound recommendations.

The value of this case is therefore analytic rather than representative. Extreme cases can expose mechanisms less visible under moderate conditions~\cite{flyvbjerg2004five}. We characterize these dynamics through two related patterns: \emph{infrastructural dependency capture}, where access to digital resources becomes entangled with intermediary control, and \emph{compound threat-attribution uncertainty}, where overlapping threat sources and access paths make exposure difficult to trace. Together, they show how constraints often studied separately can compound and undermine assumptions about identity, agency, infrastructure access, and epistemic support. We use these findings to derive implications for training, support infrastructures, platform design, and safer research with undocumented at-risk users.

\subsection{Compound Threats, Attribution Uncertainty, and Infrastructural Dependency}

Usable security often assumes that adversaries can be distinguished, threats attributed, and protective behaviors matched to relatively stable threat models. Our findings show the limits of this assumption. For participants, DPRK transnational pressure, Chinese digital infrastructure, employers, brokers, platforms, carriers, community members, and family-based leverage were connected through shared devices, dependencies, and information flows, producing a compound threat context rather than separable sources of risk.

Importantly, these threats were not simply additive: one could create the conditions through which another became more consequential. Identity-related threats and dependency-based threats often co-occurred: identity requirements within Chinese digital infrastructure could force undocumented users to rely on employers or brokers for SIM cards, accounts, or devices, converting infrastructural exclusion into intermediary control over communication and information flows. Dependency-based threats and exposure-based threats likewise tended to co-occur: a device or account registered under an employer's or broker's identity could remain linked to them even after the relationship ended, so that losing housing or employment could simultaneously cut off communication access and leave prior activity exposed to whoever inherited control of that account. Compound threat therefore arises not from assuming collaboration among adversaries, but from distinct threats intersecting through shared identities, devices, accounts, and dependency relationships.

This extends prior research on at-risk users. Refugees may face state institutions, smugglers, service providers, or hostile community members~\cite{simko2018computer,tran2024security}; journalists may face state actors or hostile sources~\cite{mcgregor2017weakest}; and IPV survivors may face abusers exploiting devices and accounts~\cite{freed2017digital,matthews2017stories}. Our setting further shows how multiple threats can converge through the same infrastructure and dependency relationships. Prior research on NKDs resettled in South Korea offers a useful comparison of how digital adaptation unfolds at a later stage~\cite{noh2024starting,noh2025bridging}. Its participants were learning unfamiliar technologies, using social media, and deciding how to present themselves while navigating stigma and social comparison. The NKDs in our study faced an earlier problem: gaining and maintaining access at all. Phones, SIM cards, accounts, and assistance could depend on employers, brokers, or borrowed identities. Decisions about disclosure, contacts, and channel use were therefore entangled with control over the underlying device or account. Informal support carried the same tension, since those who enabled access could also create exposure or dependency. Identity operated differently as well. After resettlement, the challenge involved disclosure and belonging; in China, participants lacked a legal identity they could safely use for identity-linked services.

The two stages are nevertheless connected. Noh et al.~\cite{noh2024starting} report that some resettled NKDs continued to contact family members in North Korea through broker-mediated channels marked by poor connectivity, high cost, and detection risk. Read alongside our participants' concerns about consequences for family members, this suggests that transnational family-related risk may outlast changes in legal status and digital access. Our data do not track that transition, however. Establishing which other practices persist, change, or become unnecessary would require research across displacement and resettlement.

Identity infrastructure was central to this convergence: for undocumented participants, it functioned simultaneously as an infrastructure of access, exclusion, exposure, and dependency. Without recognized legal identity, digital participation could require borrowed identities, employer-registered SIM cards, brokered accounts, shared devices, or phones purchased under another person's name. These arrangements enabled communication, work, and everyday participation while giving intermediaries control over registration, recovery, inspection, or continued access. This extends work on device control and third-party dependency~\cite{freed2017digital,matthews2017stories,tseng2022care,leitao2019anticipating}. Unlike settings where a third party gains control over an existing resource, here the intermediary could be the condition of access to that resource in the first place. We describe this pattern as \emph{infrastructural dependency capture}: actors enabling digital access can simultaneously gain durable control over it, while exiting the relationship may threaten livelihood, shelter, mobility, or legal safety. Account and device ownership should therefore be understood relationally: users may physically possess a device while another actor controls its registered identity, recovery channel, or authority to inspect it.

These dependencies created overlapping exposure paths. A phone registered by an employer or broker could simultaneously be subject to intermediary inspection, platform data retention, and DPRK-related monitoring of family communication. When participants suspected exposure, employer access, platform retention, DPRK-side interception, or other connected pathways could all be plausible. We describe this as \emph{compound threat-attribution uncertainty}: uncertainty produced not simply by limited technical knowledge, but by multiple plausible threat sources converging on shared infrastructure. Even technically informed users may therefore struggle to identify how exposure occurred, recalibrate risk, or determine which protection addresses which threat. Interventions targeting a single condition may consequently have limited effects while these infrastructural relationships remain coupled.

These dynamics also explain why individual recommendations such as changing passwords, obtaining new numbers, installing secure apps, or separating accounts may be insufficient: such advice presupposes device and account autonomy that may not exist. Threat models and security support for such settings should therefore trace harm pathways, infrastructures, dependencies, and users' actual control over digital resources rather than treating adversaries or account ownership in isolation.

\subsection{Schema Transfer and the Limits of Protective Practices}

This pattern extends prior work showing that security mental models are partial and situated~\cite{wash2010folk,wash2015too,krombholz2019if}. More specifically, our findings show how surveillance schemas formed in non-digital settings structured which digital threats participants could recognize. DPRK surveillance experience made interpersonal and visible risks salient, but provided fewer analogies for metadata, server-side retention, device linkage, background collection, and other less visible digital traces.

This distinction changes the intervention problem. Rather than treating participants as simply lacking digital knowledge, security support should bridge existing surveillance schemas to less visible forms of digital traceability. Schema mismatch may also broaden suspicion under ambiguous exposure, making specific technical access pathways harder to distinguish and reinforcing compound threat-attribution uncertainty.

This perspective also reframes apparently miscalibrated protective practices. Practices such as borrowing phones, cycling accounts, deleting local records, or disabling Wi-Fi had narrower protective scope than participants sometimes expected, but represented locally rational attempts to regain agency under uncertainty. This extends prior work on security advice~\cite{herley2009so,ion2015no,redmiles2020comprehensive,hielscher2023employees}: guidance must address not only what a practice protects against, but whether users have the infrastructural control necessary to enact it.

\subsection{Habituation Without Warnings: Adaptive and Erosive Trajectories}

Prior habituation research examines repeated explicit warnings and declining attention~\cite{vance2018tuning,anderson2015polymorphic,felt2012android}. Our findings extend this literature to chronic, largely invisible risk without warning prompts or immediate adversarial feedback. Habituation nevertheless bifurcated: \emph{adaptive habituation} followed concrete knowledge updates and more precise protection, whereas \emph{erosive habituation} followed the prolonged absence of visible adverse events without corresponding improvements in threat understanding. For some participants, situated knowledge developed through prior persecution appeared to sustain vigilance despite repeated exposure. From an assets-based perspective, the asset was this knowledge, not the persecution itself~\cite{wongvillacres2020culture,wongvillacres2022elevating}.

This distinction suggests that habituation is not simply declining vigilance. The absence of visible harm can itself become a feedback signal: when accompanied by credible knowledge, reduced fear can represent adaptive calibration; when time and silence are the only signals, it can produce erosive relaxation. This problem is particularly consequential under compound threats because trace production and harm may be temporally decoupled: an action may produce no immediate consequence yet become consequential through subsequent device inspection or, where available, service- or network-level records. The appropriate goal is therefore calibrated confidence rather than sustained fear.

\subsection{Community Knowledge Networks, Transnational Repression, and Digital Abstinence}

Security knowledge in this setting was not an individual resource but circulated through partners, employers, brokers, peers, NGOs, and informal networks. Prior research shows that social networks can provide displaced users with critical information and security support~\cite{simko2018computer,tran2024security}; our findings show that under transnational repression these same networks can simultaneously provide protection and create exposure. Trusted contacts may enable knowledge updates, while coercive intermediaries, rumors, or compromised community channels can distort or weaponize them.

This dual role connects knowledge networks directly to habituation. Access to credible information can support adaptive recalibration, whereas weak or unreliable information leaves the absence of visible harm as one of the few available signals, facilitating erosive habituation. Digital abstinence intensifies this tension. Although prior HCI work describes non-use as self-regulation or refusal~\cite{baumer2013limiting,schoenebeck2014giving,satchell2009beyond}, here abstinence simultaneously reduced digital exposure and restricted access to aid, warnings, community support, and security knowledge.

Community networks should therefore be understood as \emph{dual-use security infrastructure}. Under compound threats, the same relationship or channel may provide access, protection, surveillance, coercion, or misinformation; family pressure and community-mediated malware illustrate how adversaries can exploit precisely the relationships users trust~\cite{moss2022going}. Security interventions should consequently evaluate not only the content of advice but also its provenance, delivery channel, and susceptibility to capture.

\subsection{Design and Methodological Implications}

\paragraph{\textbf{Implication 1: Bridge existing surveillance schemas to digital traceability}}

Participants described surveillance partly in interpersonal terms: who might report them, inspect their phones, or share information. Training can begin with these concerns and explain less visible digital traces, such as metadata, account linkage, device identifiers, and server-side records. Because users' threat models shape responses to security advice~\cite{wash2010folk}, explanations should build on participants' existing surveillance knowledge rather than assume a generic model centered on viruses or hackers.

\paragraph{\textbf{Implication 2: Make security advice efficacy-specific and agency-aware}}

Security advice should state what a practice protects against, leaves exposed, and requires users to control. Clearing local records may remove handset evidence while leaving copies on other devices, backups, or service and network logs. Similarly, changing accounts offers limited protection when phone numbers, device identifiers, or social ties remain linkable. Generic security advice can cost more than it delivers~\cite{herley2009so,ion2015no}. Here, some recommendations also assumed control over devices, accounts, identity documents, and communication channels that participants lacked.

\paragraph{\textbf{Implication 3: Treat reachability as prior to content}}

Before deciding what information to provide, support organizations must consider whether users can safely receive it. Digital abstinence can make conventional outreach inaccessible, while community channels can expose users through relationships enabling contact. NGOs should establish safe, verifiable contact paths, avoid centralized lists of at-risk individuals, and minimize required disclosure of identity, location, and social ties. Digital non-use has often been studied as self-regulation or refusal~\cite{baumer2013limiting,schoenebeck2014giving,satchell2009beyond}; here, it also shaped whether assistance reached participants.

\paragraph{\textbf{Implication 4: Reduce dependence on identity-bound infrastructure}}

Possessing a device or using an account did not necessarily mean controlling it. This divergence, which we describe as infrastructural dependency capture (\S\ref{subsubsec:identity-dependency}), makes identity verification, phone-number binding, account recovery, and shared-device access potential risks. Where legally and practically feasible, platforms and support organizations should minimize identity collection, provide low-data or pseudonymous access, and support communication channels independent of employers or brokers.

\paragraph{\textbf{Implication 5: Make research transparency proportional to participant risk}}

Research transparency should not require details that make participants easier to identify. For undocumented at-risk users, raw transcripts, precise locations, recruitment intermediaries, and operational details can create foreseeable harm. Researchers can support scholarly evaluation by documenting sampling, analysis, and evidentiary decisions while withholding information that could enable re-identification or adversary learning~\cite{bellini2024sok}.

For undocumented at-risk users, usability depends on whether recommendations can be followed without increasing legal exposure, social visibility, or intermediary dependence, not simply whether instructions are understandable.

\subsection{Limitations}

This study has several limitations. First, only two participants (P14 and P19) showed clear adaptive habituation, leaving unclear whether trusted partners and NGO contact are unique calibration pathways or whether other mechanisms yield similar outcomes. Second, all participants resided in Northeast China, where ethnic Korean communities and surveillance conditions may differ elsewhere; our findings identify regional mechanisms rather than characterize all NKDs in China. Third, our habituation account relies on retrospective narratives and residence-duration comparisons rather than longitudinal tracking, risking memory reconstruction and confounding with gender, age, work arrangements, or prior digital exposure. Fourth, we did not forensically verify reported surveillance incidents; threatening messages or suspected monitoring, though technically and politically plausible, cannot be independently confirmed. We therefore distinguish participant-reported experiences, documented infrastructural mechanisms, and our own threat-relative interpretation of protective scope. These assessments characterize plausible protection under stated threat models, not whether any platform or actor exercised every technically possible capability. Fifth, ethically necessary snowball sampling may underrepresent isolated, deeply hidden, or intermediary-distrustful individuals, potentially underestimating digital abstinence and restricted access to trustworthy security knowledge. Sixth, telephone or Tencent Meeting interviews generated network- or service-level metadata. Tencent Meeting reports collecting IP, device, unique-identifier, operation-log, and meeting-timing information. Participant-selected modality avoided unfamiliar tools but could not eliminate platform-level metadata exposure, which we could not independently audit~\footnote{\url{https://meeting.tencent.com/privacy-policy-en.html}}. Finally, participants may have withheld operational details for safety, and Korean-to-English translation may have lost affective nuance despite team review. These limitations reflect methodological and ethical constraints of research with a highly vulnerable population.

\section{AI Usage Disclosure}
Generative AI tools were used solely for copyediting purposes, including checking grammar and improving phrasing. 

\section*{Acknowledgments}

Lionel Z. Wang is the corresponding author. The authors thank all participants for their valuable contributions to this work.


\bibliographystyle{IEEEtran}
\bibliography{custom}




\clearpage
\onecolumn 
\appendices

\section{Interview Protocol}
Table~\ref{tab:interview-protocol} presents a condensed version of the semi-structured interview protocol. The protocol was organized to move from less sensitive background questions toward more sensitive topics concerning digital security, surveillance perception, and coping practices. This sequencing was designed to reduce participant discomfort and to allow interviewers to adapt follow-up questions according to each participant's safety situation, level of digital familiarity, and willingness to discuss particular experiences.


\begin{table}[H]
\centering
\caption{Condensed semi-structured interview protocol}
\label{tab:interview-protocol}
\small
\setlength{\tabcolsep}{4pt}
\renewcommand{\arraystretch}{0.95}
\begin{tabularx}{\textwidth}{@{}p{0.30\textwidth}X@{}}
\toprule
\textbf{Core Question} & \textbf{Key Probes} \\
\midrule

\multicolumn{2}{@{}l}{\textbf{A. Perceived identification and tracking risks}} \\
\midrule

If you think your identity or background could be discovered, who might find out, and how?
& Most concerning possibility; different parties and methods; scenario involving posting a self-photo online. \\

How do you think authorities or others might track a person's activities?
& Concerning methods; sources of knowledge, including personal experience, others' accounts, or online sources; perceived reliability. \\

What do you pay attention to when using phones or the Internet?
& Identification practices; what is shared; communication choices; whether practices were self-developed or guided; perceived effectiveness and uncertainty. \\

Have you ever thought something was safe to do, and later changed your view---or the other way around?
& Example; reason for changed understanding; behavioral consequences. \\

\midrule
\multicolumn{2}{@{}l}{\textbf{B. Privacy practices and communication}} \\
\midrule

When contacting family members or others from North Korea, what methods do you use?
& Reasons for choosing methods; awareness or use of WeChat, Telegram, KakaoTalk, or other tools; perceived suitability; awareness of end-to-end encryption. \\

Do you share photos or other content on social media?
& Appropriateness of sharing; decisions not to post; location metadata; screenshots or forwarding by others. \\

\midrule
\multicolumn{2}{@{}l}{\textbf{C. Privacy knowledge and misconceptions}} \\
\midrule

Where did you learn what you know about phone and Internet privacy?
& Friends, employers, support organizations, or self-learning; whether anyone taught digital protection; perceived reliability. \\

Have you come across terms such as VPN, encryption, or anonymity?
& Familiarity and use; learning sources; usability; awareness of legal restrictions on VPN use in China; effects on behavior. \\

If someone told you that your phone could still be located even when Wi-Fi is turned off, how would you respond?
& Plausibility; possible behavioral changes; prior misunderstandings about technology; recoverability of deleted pictures; translation-app data storage; perceived safety of Telegram. \\

How would you describe the likelihood of someone in your situation being identified through digital means in China?
& Basis for risk assessment; known exposure incidents; perceived adequacy of current practices. \\

\midrule
\multicolumn{2}{@{}l}{\textbf{D. Changes over time and closing}} \\
\midrule

How has your awareness of these risks changed since you first arrived in China?
& Increased or decreased concern; timing and causes of change. \\

Have your phone-use habits changed since you first arrived?
& Former versus current practices; whether changes reflect external conditions or personal adjustment. \\

Among people you know who have been in China for different lengths of time, have you noticed differences in how they think about these issues?
& Differences between recent arrivals and longer-term residents. \\

Has anything caused you to reconsider your sense of risk or safety?
& Specific event or gradual development; resulting behavioral changes. \\

Were there any questions during today's interview that you found uncomfortable or would prefer not to have answered?
& Additional information the participant wants to share. \\

If you have any thoughts later or would like to follow up, do you know how to reach us?
& Provide safe contact information and inform the participant of available support resources. \\

\bottomrule
\end{tabularx}
\end{table}

\twocolumn 
\section{Meta-Review}
The following meta-review was prepared by the program committee for the 2027 IEEE Symposium on Security and Privacy (S\&P) as part of the review process as detailed in the call for papers.
\subsection{Summary}
The paper presents insights from interviews with North Korean Defectors in Northeast China. The paper highlights that this is a particular at-risk population with specific challenges arising from continuing pursuit by the North Korean regime, the legal precarity and the surveillance infrastructure within China, and the challenges arising from transposing models of physical surveillance with limited exposure to the Internet and smartphones before defection. The paper proposes a number of recommendations on this basis.
\subsection{Scientific Contributions}
\begin{itemize}
\item Provides a Valuable Step Forward in an Established Field
\end{itemize}
\subsection{Reasons for Acceptance}
\begin{enumerate}
\item The paper studies a population that is highly at-risk, hard to reach and contextually unique given the challenges. It covers a population with complex digital security and privacy needs and outlines the implications of those needs.
\item The paper communicates what the research community can learn from this context regarding methodological considerations when conducting studies with such high-risk populations.
\item Studying NKDs who remain undocumented in China adds an important context that is different from prior studies of at-risk populations and NKDs after resettlement in South Korea.
\end{enumerate}

\end{document}